\documentclass[a4paper]{jpconf}
\usepackage{graphicx}
\usepackage{epsf}
\usepackage{color}
\usepackage{subfigure}

\begin{document}
\title{Setting up a STAR Tier\,2 Site at Golias/Prague Farm}

\author{Petr Chaloupka, Pavel Jakl, Jan Kapit\'{a}n and Michal Zerola}
\address{Nuclear Physics Institute\\
  Academy of Sciences of the Czech Republic\\
  Prague, Czech Republic}
\ead{petrchal@rcf.bnl.gov}

\author{J\'{e}r\^{o}me Lauret for the STAR collaboration}
\address{Brookhaven National Laboratory\\
  Upton, USA}
\ead{jlauret@bnl.gov}

\begin{abstract}
High Energy Nuclear Physics (HENP) collaborations' experience show
that the computing resources available at a single site are often
neither sufficient nor satisfy the need of remote collaborators eager to
carry their analysis in the fastest and most convenient way.
From latencies in the network connectivity to the lack of interactivity, 
work at distant computing centers is often inefficient.
Having fully functional software stack on local resources is a strong
enabler of science opportunities for any local group who can afford
the time investment. The situation becomes more complex as vast amount
of data are often needed to perform meaningful analysis.

Prague's heavy-ions group participating in STAR experiment at RHIC has
been a strong advocate of local computing as the most efficient means of
data processing and physics analyses. To create an environment where
science can freely thrive, a Tier~2 computing center was set up at a
Regional Computing Center for Particle Physics called "Golias". It is the
biggest farm in the Czech Republic fully dedicated for particle
physics experiments.

We report on our experience in setting up a fully functional Tier~2 center
leveraging the minimal locally available human and financial resources. We
discuss the solutions chosen to address storage space and analysis
issues and the impact on the farms overall functionality.
This includes a locally built STAR analysis framework, integration with
a local DPM system (a cost effective storage solution), the influence of the availability and
quality of the network connection to Tier~0 via a dedicated CESNET/ESnet link
and the development of light-weight yet fully automated data transfer
tools allowing the movement of entire datasets from BNL (Tier~0) to Golias
(Tier~2). We will summarize the impact of the gained computing
performance on the efficiency of the local
physics group at offline physics analysis and show the feasibility of such
a solution that can used by other groups as well.
\end{abstract}

\section{Overview of STAR experiment computing}
The STAR (Solenoidal Tracker At RHIC)\cite{haris-STAR,star_nim} experiment at
Brookhaven National Laboratory (BNL) is a collaboration of over $500$ scientists
from 55 institutions in 12 countries.
It is the largest high energy nuclear experiment in the world currently taking data.
One of RHIC's goals is to study nuclear matter at extreme
conditions (pressure and temperature) similar to those that existed in the
very early universe by colliding heavy ion nuclei\cite{star_whitepaper}.
In parallel with this program, with increasing importance, there is also a unique
proton-proton program to study proton spin structure\cite{pp_rhic}.
To achieve these ambitious tasks, STAR collected large volumes
amount of data since its first run in year 2000 and the rate at which data are
accumulated is increasing.
Over the 9 years of running, STAR has collected $~7$PB of physics data in total.

\begin{figure}
\begin{center}
\includegraphics[width=8cm]{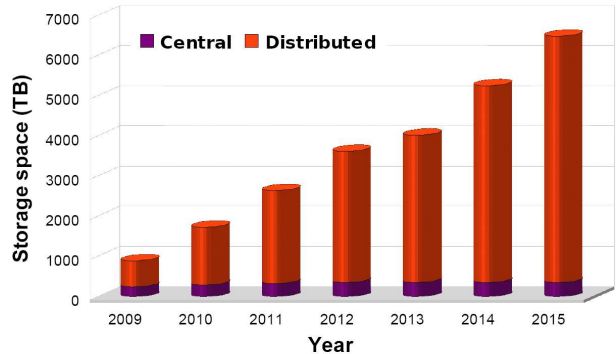}
\end{center}
\caption{\label{fig:storage2}
Projected disk storage capacity required by STAR software together with
the relative portion of space allocated
to central (NFS like) and distributed (cheap disks attached to
processing nodes) storage as a function of year.
}
\end{figure}

In order to remain on the forefront of nuclear research, the STAR
and RHIC are undergoing upgrades heading toward high-luminosity beams and detectors focused
to harvest the fruits of the most challenging physics topics\cite{decadal_plan}.
The STAR  upgrades will include new detector
instrumentation targeted to enhance STAR's acceptance,
particle identification capability, and effective sampling of luminosity.
In particular, new front end electronics for the STAR Time Projection Chamber (TPC)
and data acquisition
readout electronics will lead to an order of magnitude increase in the bandwidth for
acquisition of minimum bias data and operation for rare triggers with nearly zero dead
time, increasing the effective luminosity.
These improvements will lead to a dramatic increase in the volume of data accessible
for physics research with these probes by 60-70\% in heavy ion and proton-proton interactions.

To capitalize on these investments, it is essential that the computing
capability of the STAR experiment, now and into the future, be strategically
positioned to receive and analyze the flood of data which the upgraded
STAR detector will collect.
Part of the plan to accomplish
smooth delivery of physics results is to leverage the use of Tier~2
computing centers for user's analyses.
This approach utilizes locally
available storage and computing power and increases the productivity of
scientific work at a favorable hardware and human resource cost
affordable by the local group.

\section{STAR computing model}

Located at BNL, the RHIC Computing Facility (RCF) at BNL serves as a Tier~0 center for the STAR
experiment.
Its primary mission is to provide storage resources for mission critical data as well as at
least one pass data reconstruction and some analysis bandwidth at a lower priority.
The collected data are stored in the primary mass storage system and reconstructed later
at the RCF.
The final physics-ready data is then made available to users.
At the present time the RCF has 350 computing nodes with 1500 CPUs and 500TB of disk space dedicated to STAR only.
The STAR's needs for CPU and storage space (Figure~\ref{fig:storage2}) are expected to steadily grow 
hand in hand with the planned increase of the volume of collected data.

Until recently STAR had a single Tier~1 center: the Parallel Distributed Systems
Facility (PDSF) located at the National Energy Research Supercomputing Center (NERSC) at
LBNL.
PDSF's main purpose is to provide STAR with supplemental user analysis
cycles and simulation (aka embedding) cycles.
In addition STAR will soon open a
second Tier~1 center at the Korean Institute of Science and Technology (KISTI) in Daejon,
South Korea, a center which has the capacity to serve as a regional data redistribution hub
for STAR collaborators in Asia and provide resource boost for data production and
simulations.

\begin{figure}
\begin{center}
\includegraphics[width=7cm]{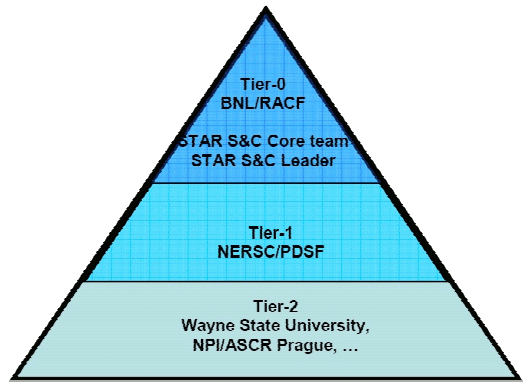}
\end{center}
\caption{\label{fig:triangle}Current structure of the STAR Software and Computing (S\&C) effort.
In addition, a Tier-1 center in South Korea is in development.}
\end{figure}

Although initially STAR mainly relied on centralized user analysis facilities (BNL and
PDSF) to provide the bulk of the analysis power for STAR collaborators and scientists, the
steadily growing amount of experimental data taken and increased demand for storage and
processing power required to reconstruct data caused the computing model to naturally
evolve toward multi-Tier structure, as shown in the Figure~\ref{fig:triangle}.
As analyses become more complex, individual users compete for available resources along
with simulation and reconstruction processes at the main Tier~0 and Tier~1 centers.
Since there can be no analysis without reconstructed data, the requirement to allocate
resources for reconstructing data in a timely fashion often takes precedence over user analyses.
A squeeze-out at Tier~0 and Tier~1 centers
motivates local groups with their own resources to set up Tier~2 centers.
The STAR software and computing plan\cite{decadal_plan} partly relies on this migration of analysis toward
Tier~2 centers serving as a buffer to resource constraints at primary facilities.
Furthermore, it has been empirically observed in STAR that productivity seem to increase
as local resources are utilized making such a computationally distributed model even more attractive.

\section{Paradigm of local computing: bringing data to users}
In world wide collaborations such as STAR, users connect to the main computing Tiers
from distant locations, often over unreliable and slow networks.
This combined with having to share computing resources and disk storage with
reconstruction and simulation jobs renders their work inefficient.
Especially, interactivity necessary to carry out analysis is hindered as well as the research objective.
These problems are even more apparent before major conferences when more computing
resources are required by users.
However, at many places computing and storage resources are available at smaller local
computing farms that can be utilized as Tier~2 centers for data processing and physics
analyses of the local physics groups after some basic prerequisites are fulfilled and the
STAR software stack is deployed.
Although the resources available at Tier~2 sites are modest, these sites strongly leverage
the capabilities at the STAR Tier~0 and Tier~1 sites with respect to scientific productivity
due to the involvement of scientist end-users in data manipulation and handling.
Such an approach allows running analyses and simulation on locally stored copies of data,
either full data sets or specifically pre-filtered subsets.
The short distance between the user and the computing center makes fast interactive work viable
and quick retrieval of computed results becomes possible thus shortening overall
turn-around time for scientific papers.

The advantage of a Tier~2 center is not only in increase of the efficiency of scientific
work, but also in cost effectiveness.
The setup of Tier~2 centers relies on building on already existing infrastructure, such as
a network or cooling system, and sharing maintenance costs with other users of the computing
farm.
As the users are expected to work predominantly with copies of data and performance (number of users)
is much lower than what is required for a Tier~0
facility, it is possible to use cheap retail class storage components such as SATA disk
arrays to store the data, and further decrease the total computing costs.
Example of above mentioned approach is a setup of a STAR Tier~2 site at the already existing
infrastructure of Golias farm in Prague.

\section{The Regional Computing Center for Particle Physics in Prague - the Golias farm}

\begin{figure}[h]
\begin{center}
\includegraphics[width=7cm]{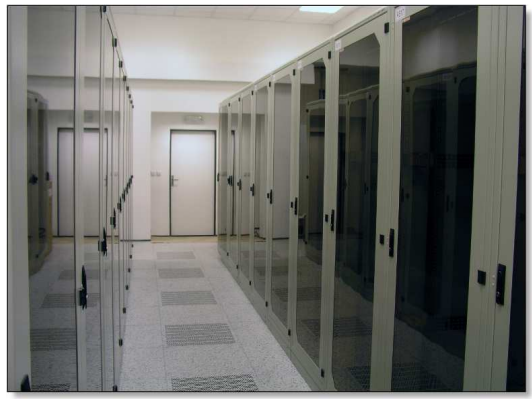}
\end{center}
\caption{\label{fig:farma}Regional Computing Center for Particle Physics, Prague}
\end{figure}

The Regional Computing Center for Particle Physics (Figure~\ref{fig:farma}), named Golias, is the biggest site in
the Czech Republic.
It is mainly dedicated to simulations and data processing for particle physics
experiments.
The farm resides in the Institute of Physics of the Academy of Sciences of the Czech
Republic and provides computing and storage services for particle physics experiments as
well as for solid state physics and astrophysics communities.
The computing and storage resources consisted of of about 450 CPUs and 50 TB of disc
space in the end of 2008 and will be gradually expanded to about 1550 CPUs/3,3 MSI2k and
216 TB of disc space in 2009.

\begin{figure}[h]
\begin{center}
\includegraphics[width=0.98\textwidth]{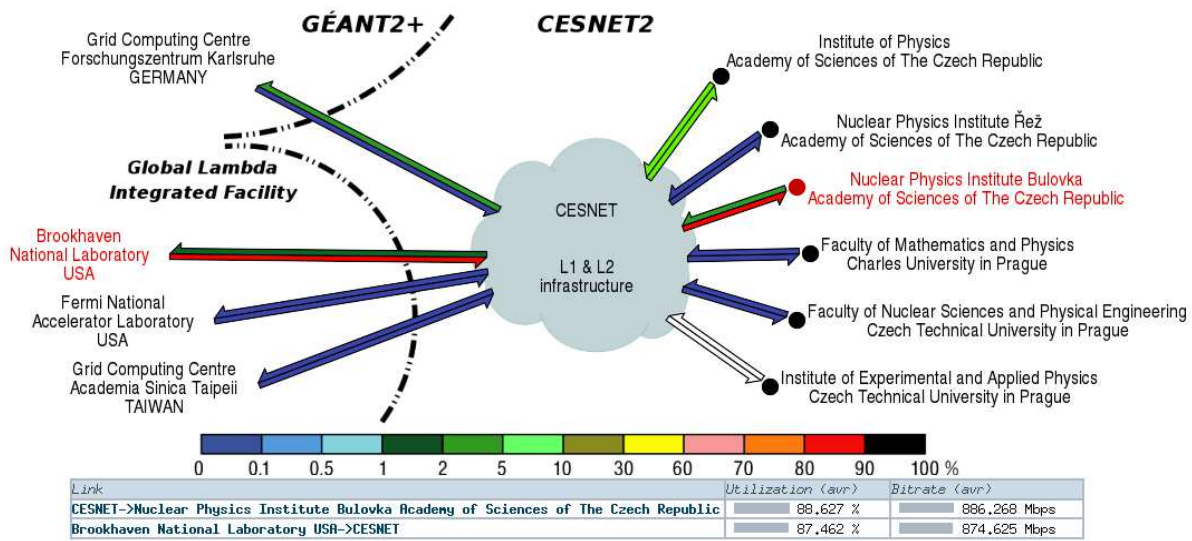}
\end{center}
\label{fig:cesnet}
\caption{Online monitoring of network connectivity provided by CESNET.}
\end{figure}

The network connectivity of the farm is maintained by CESNET (Czech Education and Scientific
NETwork)\cite{cesnet}, a non-profit association founded by universities and the Academy of
Sciences of the Czech Republic. Its aim is to conduct research in advanced network
technologies and provide network connection to educational and research institutions.
Excellent network connectivity with other institutions, both national and
worldwide (1 or 10 Gb/s) (Figure~\ref{fig:cesnet})
allows Golias to be integrated to other projects (LCG, EGEE, SAMGrid) through
their experiments' specific middle-ware components and offer its capacity to all qualified
members of supported projects (ATLAS, ALICE and D0).
The installed gLite grid middle-ware components include CE, SE, UI, MON box, site BDII and
have been configured with YAIM, integrated into Cfengine used for the management of the
local site changes.
Since 2005, the site is a certified Tier~2 center of the LCG project and in 2008 signed the
Memorandum of Understanding of the Worldwide LHC Computing Grid Collaboration (WLCG).
Part of the farm's capacity is available for local users from other projects such as AUGER and
STAR and spare cycles can be used by anyone, making the facility a good example of
efficient resource usage.

Job management at the Golias farm is performed using the workload management system PBSPro (current version 9.2)\cite{pbs_pro}.
Golias is in the process of evaluating alternatives such as Torque\cite{torque} and the cluster scheduler MAUI\cite{maui}.
The hardware, software and network status are extensively monitored using many standard packages including Nagios, Munin, MRTG, IPAC as well as a set of locally developed custom tools tailored to the facility's need.

\section{Setup of the local STAR computing environment}

The local ultra-relativistic heavy ion collisions group from the Nuclear Physics Institute
of the Academy of Sciences of the Czech Republic (NPI ASCR) has been participating in the
STAR experiment since its very first collisions in 2000.
It's main activities have been data analyses in the area of correlations and high
momentum physics as well as detector research and development activities.

\begin{figure}[h]
\begin{center}
\includegraphics[width=0.7\textwidth]{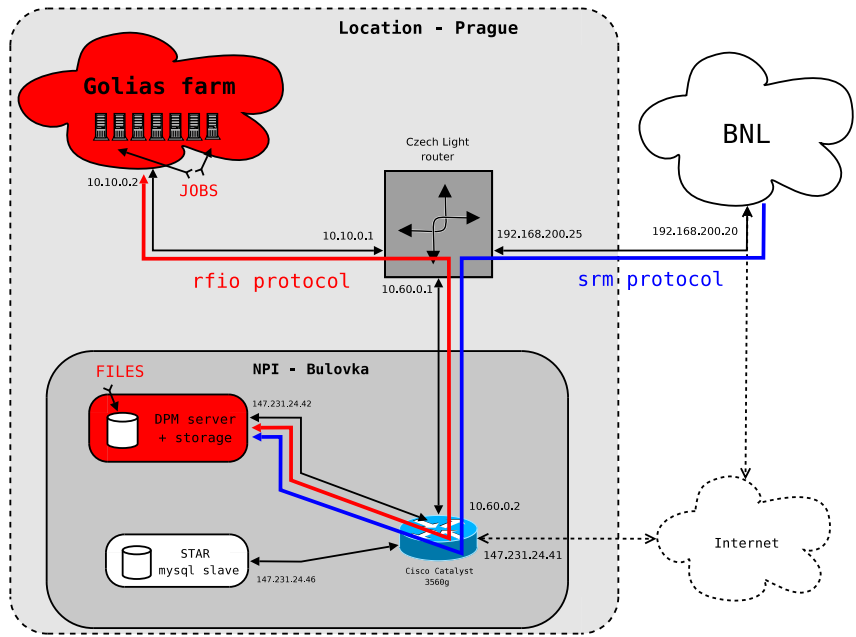}
\end{center}
\label{fig:scheme}
\caption{Schematics of Tier~2 center at Golias farm.}
\end{figure}

The road towards productive work was however paved with difficulties. In the
beginnings the group had struggled with running analyses on the major Tier centers from
its remote location given the slow connectivity over the Atlantic.
This was extremely inefficient, as then the network capabilities were an order of magnitude
slower than the ones available today.
The idea of developing local STAR computing environment has been tested since
2002 and a decisive advance was achieved by joining the Golias farm at 2005, leveraging the
knowledge acquired over the years to set up the necessary components for a Tier~2
center.
Since then, the STAR physics group has used the Golias farm for short-time and mid-scale
analyses.
Over this period of time we have tested multiple approaches and identified key components
for the successful setup of local Tier~2 computing center.
A schematic layout of the STAR Tier~2 at Golias in the configuration described
by this article is shown in the Figure~\ref{fig:scheme}.

\subsection{Deployment of STAR software framework}
First the STAR software framework must be deployed and adapted
to the specifics the of local farm.
In-situ compilation of the STAR software stack and import of user environment to emulate
the RCF is a key aspect making the work at any facility interchangeable hence transparent
(learn once, use many times).
The deployment of the STAR software stack includes a version of ROOT and a set of add-ons
customized by the STAR experiment called "root4star".
For the needs of the Golias farm, the ROOT framework was patched to support the RFIO protocol
which is used to access locally stored data, as will be explained in the next section.
To simplify the selection of data for analysis and subsequent job submission STAR has
developed the "STAR Unified Meta Scheduler" (SUMS)\cite{sums} tool and middle-ware.
Customizing SUMS for the Golias environment requires an adaptation of policies to make
use of PBSpro and Torque for scheduling, but from a user stand point, the use of SUMS and
its job syntax is identical regardless of the facility, allowing the user to simply
migrate their work from the Tier~0 to the Tier~2 center using the same job description
files.

\subsection{Data storage,transfer and aggregation strategy}
 Data management was found to be a key component of our local computing strategy.
 The overall aim is to allow physicists to easily download data sets from the RCF to
 local storage and also to provide them with a simple and fast way to connect their jobs
 running on computing nodes of the farm with the storage.
 As trivial as it may sound, at the same time one requirement was that the access to
 the data not strain internal network of the farm, as would mounted NFS.
 This combined with an additional requirement of a low purchase and maintenance cost
 resulted in the use of the SRM-based Disk Pool Manager (DPM) grid-aware solution
 developed at CERN\cite{dpm}.
 In order to keep the farm maintenance cost low DPM was actually located outside of
 the farm premises\cite{our_dpm} and is connected to Golias via a dedicated 1Gbit/s
 optical fiber link.
 The DPM offers various interfaces for data access (RFIO, XROOTD) used in the STAR
 computing framework.
 The DPM solution is very convenient for end-users since it affects their jobs in a
 minimal way and allows for easy scalability and maintenance by the administrators.
 The DPM manages about 20 TB of disk space and with sophisticated data-mover
 tools developed by the group at Prague and STAR software staff, physicists can download
 almost 1 TB of data per 24~hours from BNL.

\subsection{Network connectivity}
Fast networking available to the users of the Golias farm makes
the main difference between working remotely at major Tier centers and working locally at smaller
farm for the end-users.
As already stated above, the Golias has an excellent connection to other major
scientific institutions via CESNET.
In order to efficiently transfer data to and from the RCF, a dedicated 1Gbit/s
intercontinental link between BNL and Golias was established through cooperation
between CESNET, ESNet and BNL.
More than a year of operations show that about 1TB of data per day can be easily
transferred from BNL and written to DPM storage using SRM transfer tools.
Since DPM storage lies outside of the farm the transfer of data to running jobs at
working nodes is of a concern.
At this point a 1Gbit optic fiber link is setup between the storage facility and computing the farm.
Stress tests show that under conditions of typical physics analyses jobs run by
the group the DPM is able to supply data to $~200$ running jobs without significant slow
down of the jobs.
Above this threshold, the IO is a limiting factor which may require the consolidation of
the Local Area Network.

\subsection{Local database mirror}
Database access to the main calibration, geometry and status information database at the local farm
is a small, but very helpful addition to the local setup.
Even though we have high throughput connection at our disposal, the large distance of jobs
from main database located at the RCF caused major prolongations of running jobs due to
latencies inherent to the communication protocol.
Access to a local database mirror enables running detector simulations and embedding, if
needed, at the same speed as at the Tier~0 facility.
Since the database mirroring is a fully automatic process done through a MySQL master/slave
replication, the local database mirror is a maintenance-free addition to the Tier~2
center.

\section{Performance}

The setup of a STAR Tier~2 center at the Golias farm, in the above described configuration,
has been tested by the Prague physics group for more then a year in
conditions typical for small to mid-size physics working group.
While such a group does not always need high processing power it needs intermittent
on-demand access to computing resources.
The most common tasks performed at the Golias site are physics analyses on locally
stored, usually preprocessed, data sets on the order of $10-1000$GB.
With the available connection to the RCF facility, where primary copies of data are
stored, users are able to transfer data for their analyses with minimal time delays.
Hence, switching between analyses and data sets is possible and a wide range of studies have
been carried out locally.

Since the setup of a Tier~2 center must to be cost-effective
and simple, it has limitations and future bottlenecks are predicted.
We expect to increase the number of running jobs in future increasing the
data transfer from the DPM storage.
Our disk-array setup of DPM is currently based on a single bus/NIC which may cause a bottle
neck for data-hungry jobs in the future.
For this reason another separate storage server is being purchased to distribute
the disk-space among several loosely coupled disk-servers and balance the overall load.
Another limitation follows from the chosen low-cost solution.
Our experience shows that this solution works well at a smaller site, like ours, however
if Golias Tier~2 were to grow to the size of a regular Tier~1 center, such a solution
would be rather human resource hungry and likely inefficient.

\section{Conclusions and perspectives}
The experience with the Tier~2 center in Prague
shows  that a light-weight distributed computing  approach, concentrating on the
redistribution of data, can improve analysis viability, sustainability, and reliability
and increase the local scientific opportunity by carrying out local analyses and leveraging
local human and hardware  resources.
This approach not only relieves the load from main computing centers (BNL/RCF and
NERSC/PDSF) but also takes advantage of faster data access and shorter waiting times
for the local users.

In the near future STAR experiment has to move to grid-oriented
computing. Such a need is driven by steadily increasing data volumes expected
in the next 5-10 years.
It is thus very likely that the Golias farm as well as the
NPI ASCR local data storage will be used to process grid-submitted jobs.

\section{Acknowledgements}
The authors wish to thank the team of Golias farm for their support
of the STAR Tier~2 project. We also thank the BNL computing team for 
continuous support and development of the off-site computing facilities.
We would like to extend our gratitude to people from ESNet, and CESNET.

This work was supported in part by the Grant Agency of
the Czech Republic, Grant 202/07/0079 and Ministry of Education of the
Czech Republic, Grant LC 07048.

\end{document}